\documentclass[
reprint,
superscriptaddress,
aps,
longbibliography,
prapplied,
]{revtex4-2}

\usepackage{graphicx}% Include figure files
\usepackage{amssymb,amsmath,amsfonts, stmaryrd, latexsym}
\usepackage{mathrsfs} % math font \mathcal
\usepackage{multirow}
\usepackage{xcolor, soul}
\usepackage[top=2.5cm, bottom=2.5cm, left=2cm, right=2cm]{geometry}
\usepackage{threeparttable}

\newcommand{\image}{\hat{\mathcal I}}
\newcommand{\Is}{I_{\rm s}}
\newcommand{\zjump}{\llbracket z \rrbracket}

\begin{document}

\title{Subnanometer Accuracy of Surface Characterization by Reflected-Light \\Differential Interference Microscopy}
\author{Ka Hung Chan}
\affiliation{Department of Mechanical and Aerospace Engineering, Hong Kong University of Science and Technology, Clear Water Bay, Hong Kong}
\author{Shengwang Du}
\email{dusw@utdallas.edu}
\affiliation{Department of Physics, The University of Texas at Dallas, Richardson, Texas 75080, USA}
\author{Xian Chen}
\email{xianchen@ust.hk}
\affiliation{Department of Mechanical and Aerospace Engineering, Hong Kong University of Science and Technology, Clear Water Bay, Hong Kong}

\baselineskip20pt

\begin{abstract}
    We theorize the surface step characterization by reflected incoherent-light differential interference microscopy with consideration of the optical diffraction effect. With the integration of localization analysis, we develop a quantitative differential interference optical system, by which we demonstrate that the axial resolution of measuring surface height variation is sensitive to the shear distance between the two spatially differentiated beams. We fabricate three nanometer-size steps by photolithography, and successfully characterize their 1D height variations with $0.13$ nm$/\sqrt{\mathrm{Hz}}$ axial precision. Our result suggests that the optical differential interference microscopy can be used for real-time characterization of surface structure with a subnanometer accuracy and a large field of view,  which is greatly beneficial to the surface characterization of micro/nano-electromechanical systems. 
\end{abstract}

% \keywords{Quantitative phase differentiation, differential interference microscope, surface characterization, error analysis}

\maketitle

% \begin{linenumbers}

\section{Introduction}
Quantitative surface topography determination plays an important role in manufacturing design and control of micro/nano electromechanical systems \cite{WU201792, etchcontrol}. The surface roughness and local height variation are the essential structural parameters governing the electromechanical  responses of the micro-devices in engineering. From research and development point of view, the atomic force microscopy (AFM) is a popular quantitative characterization techniques widely exploited in this field \cite{raoufi2007surface, crozier2000thin}. The AFM uses a scan-based probe to acquire the surface profile by sensing a tiny but accurate mechanical interaction between the sample surface and the probe. As a result, the AFM provides high axial and lateral resolutions down to nanometer scales, that breaks through the optical diffraction limit \cite{raoufi2007surface, crozier2000thin, kwon2003atomic, matei2006precision}. However, the scanning speed by AFM probe is slow, corresponding to relatively small area of characterization. It usually takes minutes to complete a scan of an area of $10\times 10$ $\mu$m$^2$. It is difficult to capture the evolution of structural deformation and fracture dynamics by AFM. 

In contrast to AFM, the optical microscopy enables fast imaging of a large field of view (\emph{i.e.} $> 100 \times 100$ $\mu$m$^2$). The optical images, however, are usually not quantitative for surface topography and have lower spatial resolution as compared to AFM. There are some optical methods for quantitative 3D characterization by using optical sectioning techniques such as confocal microscopy \cite{jordan1998} and various optical interferometers \cite{wiegand1998, page2007dynamics}. For these methods, the 3D surface topography is subjected to an algorithmic reconstruction from the diffraction/interference intensities of light. Recent advances in microscope development for precise phase characterization enable applications such as the marker-free phase nanoscopy \cite{Cotte2013}, the spatial light interference microscopy \cite{Wang:11}, and the epi-illumination gradient light interference microscopy \cite{Kandel2019}. These techniques inspire a profound potential for surface topographic characterization with nanoscale accuracy by engineering the optical path gradient.

In this paper, we demonstrate subnanometer precision for determination of surface steps by a customized reflected incoherent-light differential interference microscope with variable optical differentiation parameters. The microscope we developed is similar to most differential interference contrast (DIC) microscopes \cite{deGroot:15, Shribak:13, Shribak:06, NOGUCHI20093223, Hartman:80, Nguyen-NC2017, DIC1977, DIC2001}, but we aim at quantitative measures of surface height variation instead of producing phase contrast images. We utilize the localization analysis \cite{chiu2019measuring} to precisely determine the shear distance between two orthogonally polarized light rays, by which we measure the surface topography from the phase lag between the differentiated light path \cite{zeng2019quantitative, zeng_2020jmps}. Since the phase contrast of an image is no longer the scope here, we name our optical system as Differential Interference Microscopy (DInM). Our previous works have successfully demonstrated submicron axial and lateral resolutions for measuring the full-field deformation gradient of phase-changing metals \cite{zeng_2020jmps} and the thin film buckling on soft substrate \cite{zeng_2021eml}.   

Here we push the limitation of DInM in step height measurement and demonstrate our method with variable shear distance. By considering the diffraction effect of a differential interference image, we provide a theory of reflected incoherent-light DInM and theorize an analytical expression of error for step-height measurement in terms of light differentiation parameter and wavelength. The error model guides the system design and a proper calibration, by which the experimental accuracy is much improved. Consequently, we achieve sub-nanometer precision ($0.13$ nm$/\sqrt{\mathrm{Hz}}$) for measuring the height of small steps that are fabricated on silicon wafer by the standard photolithography process. 
% \textcolor{red}{It is the best accuracy that can be achieved by using the DInM/DIC microscope.}

\section{Method and principle}\label{sec:theory}

\begin{figure}
\centering
\includegraphics[width=0.45\textwidth]{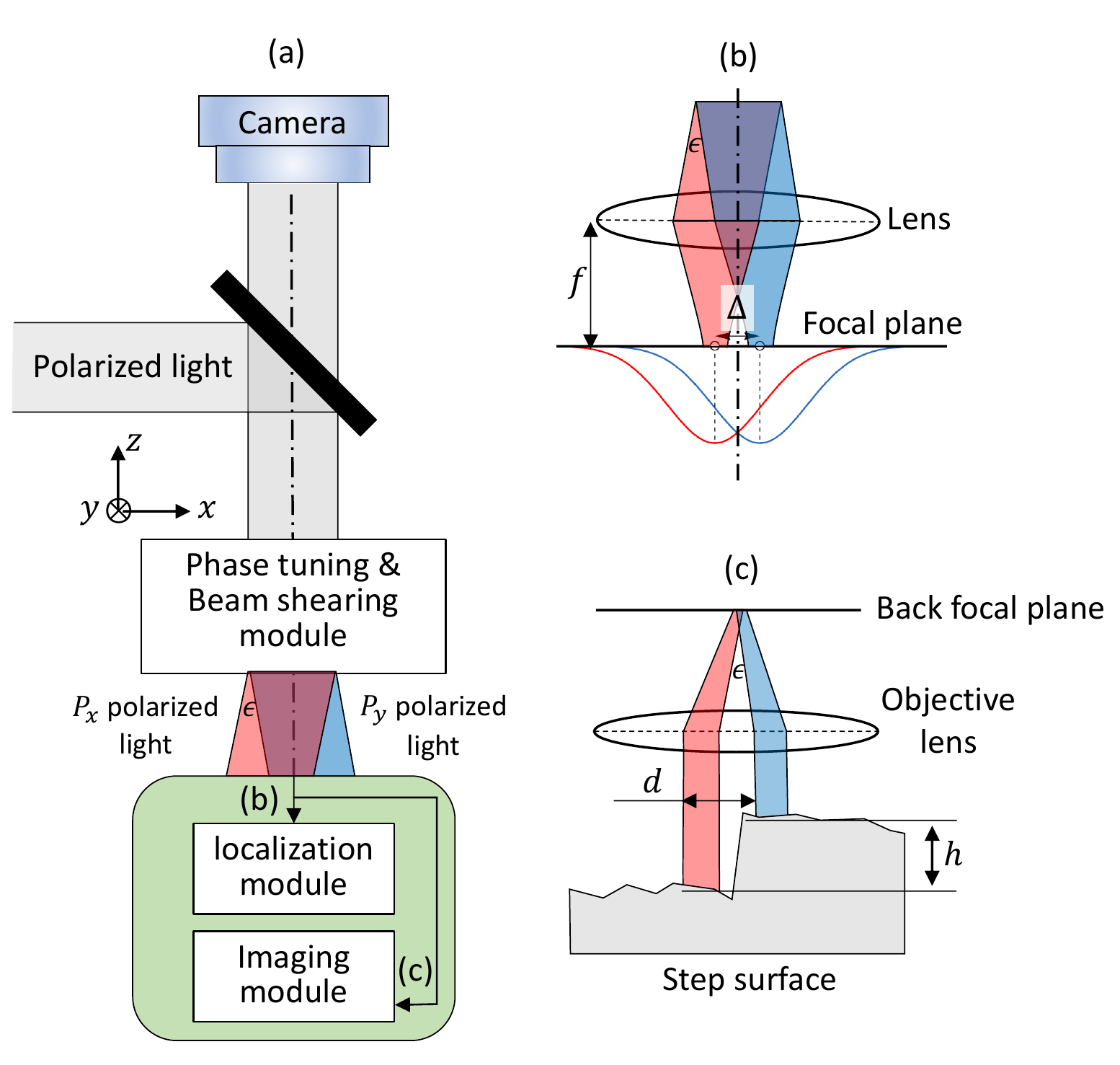}
\caption{Schematic diagram of (a) DInM optical system with integrated (b) localization module and (c) reflected-light imaging module. The two orthogonal linearly-polarized light beams (orange and blue) are spatially separated by the beam-shearing component $d$ with a tunable phase lag.}\label{fig:Setup} 
\end{figure}

The DInM optical system is comprised of three functional modules as shown in Fig.~\ref{fig:Setup}. The phase tuning and beam shearing module consists of prisms and linear phase retarders, which can spatially differentiate the propagation directions of light beams with orthogonal polarizations and gain the desirable phase lag with respective to each other. This module is similar to the Nomarski prism \cite{PMID:5361069} used for many commercial DIC microscopes, but the phase lag between the two orthogonally polarized beams can be precisely tuned \cite{zeng_2020jmps} and the beam-shear angle is variable by using the prisms with different birefringence properties. The functional modules such as the localization module and the imaging module are switchable and work independently. The localization module, as illustrated in Fig.~\ref{fig:Setup}(b), is used to measure the shear angle $\epsilon$ between the laterally separated light rays by the localization analysis. The lens with a long focal length $f$ focuses the two sheared beams into Gaussian-shape spots with a separation $\Delta$ on the front focal plane. The angle between differential light paths is determined as $\epsilon = \Delta/f$. Details of the measurement are introduced in Ref.~\cite{chiu2019measuring}. The surface step of specimen is characterized by the differential light through the imaging module as shown in Fig.~\ref{fig:Setup}(c). After passing through the objective lens with focal length $F$, the two collimated and orthogonally polarized light beams (red for $P_x$ and blue for $P_y$) are separated laterally along $x$-axis with a distance $d = \epsilon F = F \Delta/f$. 

When there exists a local height variation on the surface, the $P_x$ and $P_y$ beams are reflected by the surface at different heights and obtain an additional phase difference with respect to each other. Such a phase difference passes the information of local height variation to the imaging system so that the surface topography can be solved quantitatively \cite{zeng2019quantitative}. Here, our optical measurement is conducted by a one-dimensional light differentiation. We consider an intensity image $\mathcal I : \mathbb R^{W} \to [0, \infty]^W$ where $W$ is the number of pixels along the beam-shear direction of a two-dimensional DInM image. The intensity image $\mathcal I$ is related to the surface height variation by \cite{zeng2019quantitative}
\begin{equation}\label{eq:intensity}
\mathcal I(x) = I_0 \sin^2 \left(k \llbracket z\rrbracket(x, d) + \phi_0\right) + I_{\rm s},
\end{equation}
where $I_0$ denotes the reference intensity, $I_{\rm s}$ is the intensity caused by stray light from background, $k = \frac{2 \pi}{\lambda}$ is the wave number for wavelength $\lambda$ of incoming light and $\phi_0$ is the bias phase that can be tuned by a set of liquid crystal linear retarders. The surface height variation is defined as
\begin{equation} \label{eq:var}
    \llbracket z \rrbracket(x, d) = z(x) - z(x-d),
\end{equation}
for beam separation distance $d > 0$ (also known as beam-shear distance).
Since the reference and background intensities are constants throughout the measurement, the expression \eqref{eq:intensity} can be normalized as
\begin{equation}\label{eq:normalizedintensity}
\image(x) = \frac{\mathcal{I}(x)-I_{s}}{I_{0}}=\sin^2\left(k \llbracket z\rrbracket(x, d)+\phi_0\right).
\end{equation}

\begin{figure*}
\centering
\includegraphics[width=0.8\textwidth]{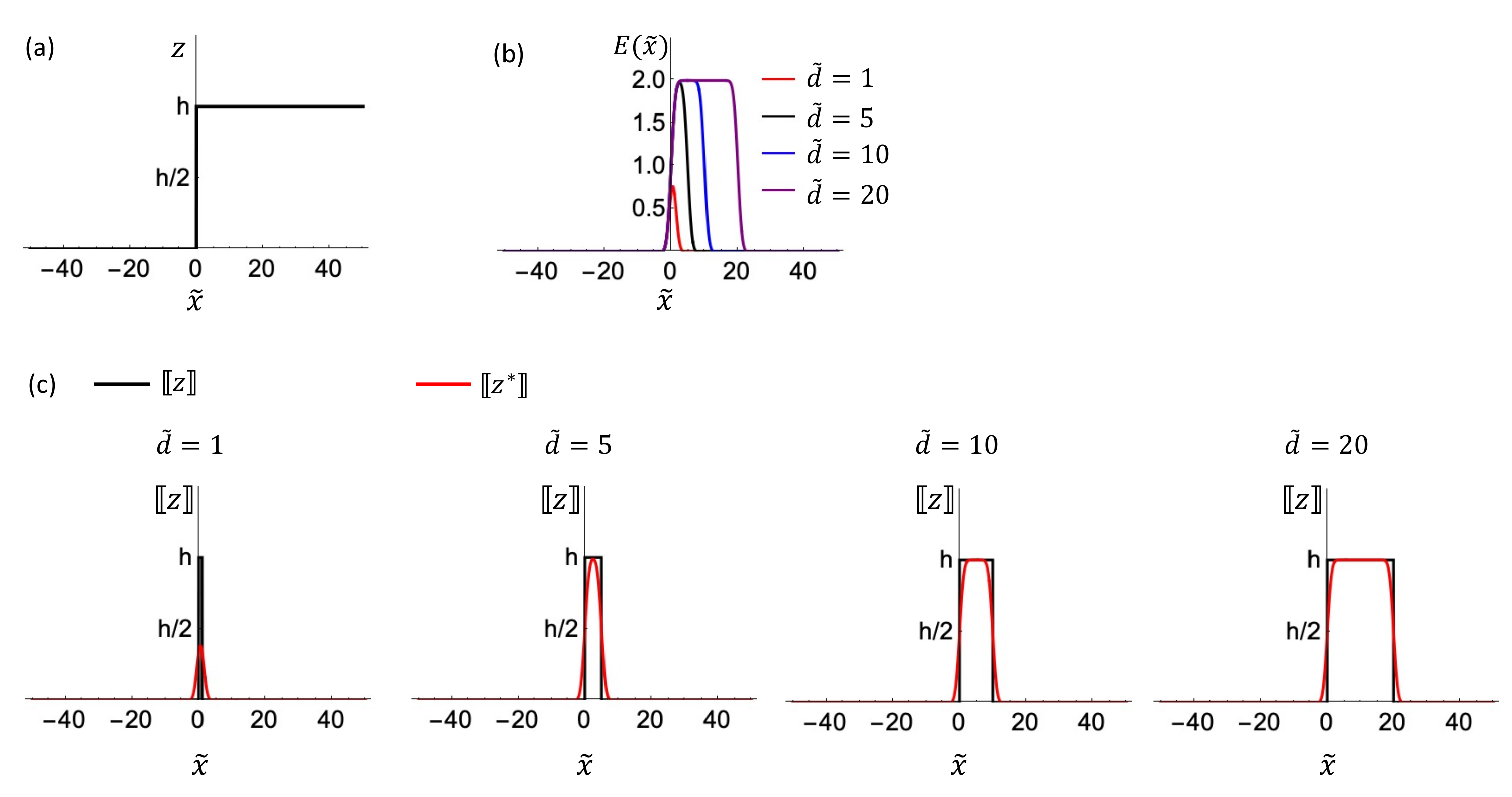}
\caption{(a) The profile of a one-dimensional step defined by Eq.\eqref{eq:step}. (b) The function $E(\tilde x)$ at various $\tilde d$. (c) Comparison between the surface height variation $\llbracket z \rrbracket$ and the modified height variation $\llbracket z^\ast \rrbracket$ under different choices of dimensionless shear parameter $\tilde d$.} \label{fig:theory} 
\end{figure*}

In principle, three independent measurements of $\mathcal{I}(x)$ by tuning the bias $\phi_{0}$ are sufficient to solve the height variation $\llbracket z\rrbracket$ on the surface.  However, in a real optical system, the acquired image is subjected to the diffraction effect by light, which results in blurriness of image \cite{krauskopf1962light}. In this work, we use LED light source without spatial coherence. While the amplitude of the light source is uniform on the transverse plane, the phase distribution is stochastic and uncorrected \cite{StatisticalOptics, fourieroptics}. Mathematically, the diffraction effect caused by finite aperture can be modeled as the convolution of the ideal image by a kernel of point spread function (PSF). Then the realistic DInM image with consideration of diffraction effect is expressed as
\begin{equation}\label{eq:convolution}
\image^*(x) =\image(x) \circledast {\rm PSF} (x)= \int \image(x') {\rm PSF}(x-x') dx',
\end{equation} 
where $\circledast$ denotes the convolution operation (see the detailed derivation in Appendix \ref{A1}). Note that we neglect the partial spatial coherence induced by the aperture of the objective lens. A more accurate but complicated derivation regarding the partial spatial coherence can be found in Ref. \cite{MehtaOE2008}, which does not affect the phase information from surface height variations. The intensity PSF kernel in our system can be well approximated by a Gaussian function: 
\begin{equation}\label{eq:PSF}
{\rm PSF}(x) = \frac{1}{\sqrt{2\pi}\sigma} {\rm exp}\left(-\frac{x^2}{2\sigma^2} \right),
\end{equation}
where the standard deviation $\sigma$ denotes the lateral resolution of the microscope. In Appendix \ref{A2}, we show that the Gaussian function in Eq. \eqref{eq:PSF} with $\sigma = 1.1 /({\rm NA}k)$ is a good approximation for the point spread function. 

As an application of Eq. \eqref{eq:convolution}, we are interested in measuring steps on surface. We define a one-dimensional step function as
\begin{equation}\label{eq:step}
    z(x) = \left\{\begin{array}{ll}
        0, & x \leq 0 \\
        h, & x > 0 
    \end{array}\right.
\end{equation}
where $h$ is the step height. According to \eqref{eq:var}, we have  
\begin{equation}\label{eq:rectangular}
\llbracket z\rrbracket(x, d) = \left\{\begin{array}{ll}
0, & x < 0 \text{ or } x > d\\
h, & 0 \leq x \leq d 
\end{array}
\right..
\end{equation}
Substitute \eqref{eq:rectangular} and \eqref{eq:PSF} into Eq. \eqref{eq:convolution}, we obtain the expression of intensity profile measured by DInM with consideration of light diffraction effect as
\begin{equation}\label{eq:normalizedintensity2}
\image^*(\tilde x) = \sin^2\phi_0 + \frac{1}{2}\sin (k h) \sin (k h + 2\phi_{0})E(\tilde x) 
\end{equation}
for dimensionless position $\tilde x = x / \sigma$. The function $E(\tilde x)$ is defined as
\begin{equation}\label{eq:errorfunc}
E(\tilde x) = {\rm erf} \left(\frac{\tilde{x}}{\sqrt{2}}\right)-{\rm erf} \left(\frac{\tilde{x}-\tilde{d}}{\sqrt{2}}\right)
\end{equation}  
where ${\rm erf}(\cdot)$ is the Gauss error function and $\tilde{d}=d/\sigma$ is a dimensionless optical shear parameter. 
% \eqref{eq:errorfunc} shows, because of the diffraction (convolution) effect, the rectangular-shape intensity profile exhibits smooth rising and falling transition length of $\sigma$.  

The discrepancy between the intensity profiles with and without convolution is sensitive to the dimensionless shear parameter $\tilde{d}$, that is the ratio of beam-shear distance to the lateral resolution. This is illustrated in Fig.~\ref{fig:theory}. For a one-dimensional step shown in Fig.~\ref{fig:theory}(a), the ideal intensity profile should be a rectangular shape with a sharp variation in height of $h$ and width of $\tilde d$. Due to the diffraction between the two beams with subtle spatial separation, the corners of the rectangle profile are smoothed out. The amount of such a smoothness is given by Eq. \eqref{eq:errorfunc}. We calculate the function $E(\tilde x)$ in Eq. \eqref{eq:errorfunc} for $\tilde d = 1, 5, 10, 20$ respectively, in Fig. \ref{fig:theory}(b). As expected, their rising and falling edges show smooth transition. As $\tilde{d}\gg 1$ (\emph{i.e.} $d\gg\sigma$), the interference between ${\rm erf} \left(\frac{\tilde{x}}{\sqrt{2}}\right)$ and ${\rm erf} \left(\frac{\tilde{x}-\tilde{d}}{\sqrt{2}}\right)$ in Eq. \eqref{eq:errorfunc} diminishes.  As $\tilde{d} \to \infty$, $E(\tilde x)$ is converged to 2 for $0 \leq \tilde x \leq \tilde d$, consequently $\image^* \to \image$. Following Eqs. \eqref{eq:normalizedintensity} and \eqref{eq:normalizedintensity2}, we define the modified height variation $\llbracket z^\ast \rrbracket$ which satisfies 
\begin{equation}\label{eq:varmodified}
\image^*(x) = \sin^2\left(k \llbracket z^*\rrbracket(x, d)+\phi_0\right).
\end{equation}
By the inverse of Eq. \eqref{eq:varmodified} at bias $\phi_0 = 0, \frac{\pi}{4}, \frac{\pi}{2}$, the modified height variation is solved as
\begin{equation}\label{eq:zjump}
    \llbracket z^\ast \rrbracket = \frac{1}{2k}\tan^{-1} \frac{\sin(2k h)E(\tilde x)}{2(1 - \sin^2(k h) E(\tilde x))}.
\end{equation}
Therefore $\llbracket z^\ast \rrbracket \to \llbracket z \rrbracket$ as $E(\tilde x) \to 2$ for $0 \leq \tilde x \leq \tilde d$. We plot both $\llbracket z \rrbracket$ and $\llbracket z^\ast \rrbracket$ corresponding to $\tilde d = 1, 5, 10, 20$ as shown in Fig.~\ref{fig:theory}(c). As $\tilde d$ increasing, the modified height variation $\llbracket z^\ast \rrbracket$ is asymptotically converged to the original value $\llbracket z\rrbracket$. Within the range of $[0, \tilde d]$ at a step, the diffraction-induced error can be computed as
\begin{equation}\label{eq:epsilon}
\epsilon = \vert \llbracket z^\ast \rrbracket/h - 1\vert.
\end{equation}
As indicated by Fig. \ref{fig:theory}, if the distance between the spatially separated beams is sufficiently large so that $\tilde d \gg 1$ for a moderate lateral resolution, the diffraction-induced convolution effect is diminished. This underlies a system design strategy to achieve the high accuracy for surface step characterization.

\begin{figure}
    \centering
    \includegraphics[width=0.42\textwidth]{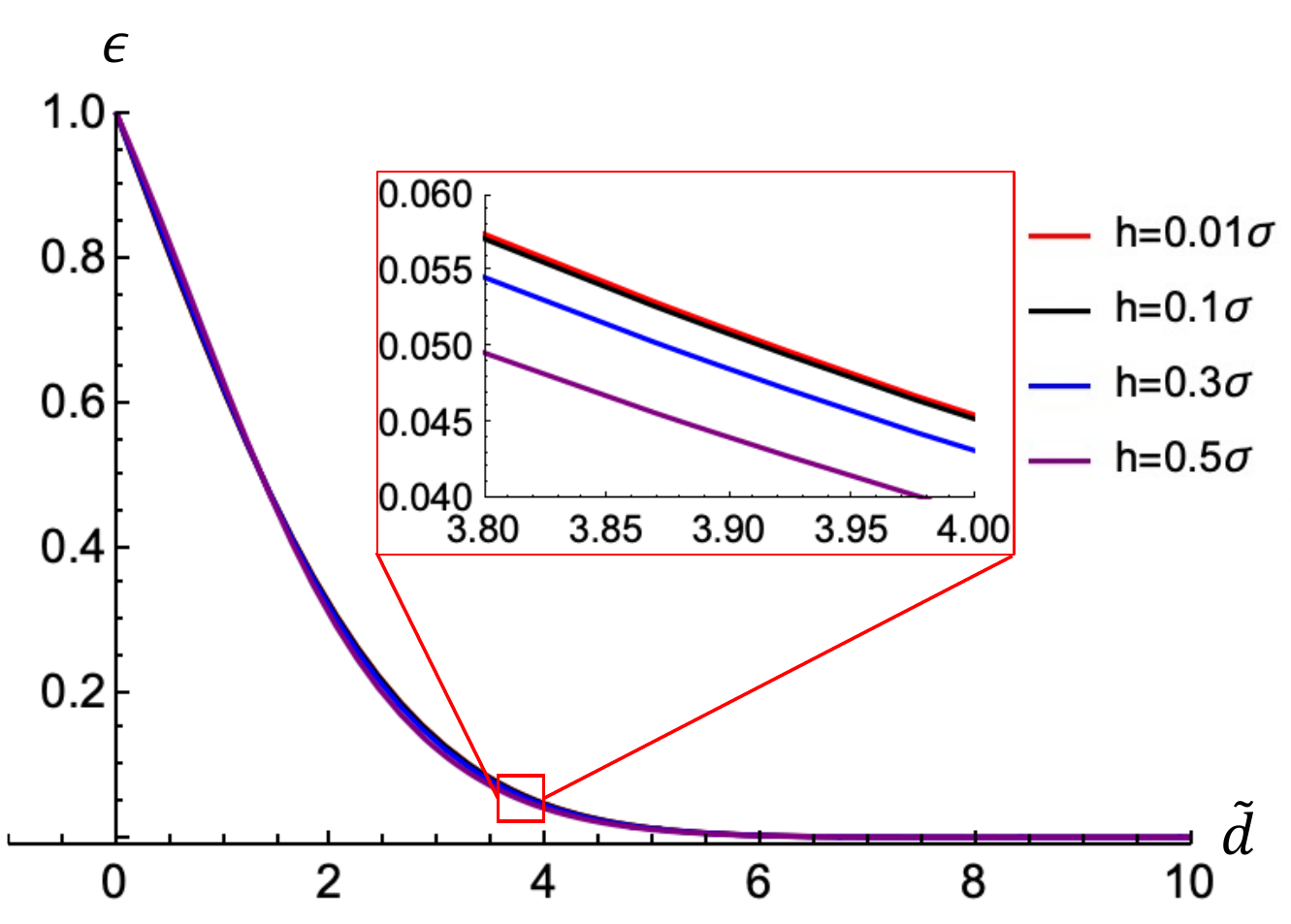}
    \caption{The diffraction-induced error versus the dimensionless shear parameter $\tilde d$ for various step height variations. Note that the step height $h$ are nondimensionalized by the lateral resolution $\sigma$.}
    \label{fig:error}
\end{figure}

Figure \ref{fig:error} shows the relationship between the diffraction-induced error and the shear parameter $\tilde d$. Although the error in Eq. \eqref{eq:epsilon} is not very sensitive to the step height $h$, we observe that the value of error increases as the decrease of the step height. An important indication from Fig. \ref{fig:error} is that the error converges as $\tilde d$ goes sufficiently large. Moreover, it suggests a selection criterion for the parameter $\tilde d$ of a differential interference microscope with desirable accuracy. 
% As the to-be-measured step height is in tens of nanometers and beyond, the nanometer accuracy can be achieved by using the $\tilde d > 4$.

Besides the systematic error (or discrepancy) caused by the diffraction effect, the axial resolution of the system is limited by a random error from the illumination and detection. By direct calculation of first-order variation of Eq. \eqref{eq:intensity}, we have
\begin{equation}
\delta \llbracket z \rrbracket = \sqrt{\left(\frac{\delta  \mathcal{I} }{k I_0}\right)^2 + \left(\frac{\delta I_0}{2 k I_0}\right)^2 + \left(\frac{\delta \phi_{0}}{k}\right)^2}. \label{eq:11}
\end{equation}
Here $\delta \llbracket z \rrbracket$ denotes the amount of linear error of the local height variation measurement from the subtle perturbations of pixelated intensity of the image, instability of light source and tuning uncertainty of optical parts for bias phase. This error can be suppressed by increasing the light source power, choosing a shorter light wavelength, and averaging over multiple measurements with a longer measurement time.

\section{Results and discussions}

We demonstrate the design strategy of the DInM optical system to characterize surface steps at nanometer scales with varying optical differential parameter $d$. The nano sized surface steps are fabricated by the photolithography process on a 4-inch poly-silicon wafer. First, we deposit a photoresist layer (HPR506) on the wafer, then pattern an array of rectangular shapes on it. The rectangular steps are finally formed by etching the unmasked region using Oxford Plasmalab 80 Plus Plasma Etcher. Steps with designed depths of $4, 7, 16$ nm are fabricated by tuning the duration for performing the photolithography process. Finally, all specimens are washed by acetone to remove the photoresist layer. The DInM optical system illurstrated in Fig. \ref{fig:Setup} uses a LED light source with central wavelength $\lambda=355$ nm. We use the Nikon TU plan fluor objective lens with ${\rm NA} = 0.3$ for imaging, which gives the lateral resolution $\sigma = 432$ nm in PSF of Eq.~\eqref{eq:PSF}. 
Different prisms are used to spatially shear the incident light into $P_x$, $P_y$ polarized light beams with shear distances in the set $\mathcal D = \{5.30, 4.40, 3.34, 2.35, 0.956, 0.423\} \mu$m respectively. The set of dimensionless optical shear parameters is calculated by 
$\tilde{d} = \frac{d}{\sigma}$, and listed as $\tilde{\mathcal D} = \{12.3, 10.2, 7.7, 5.4, 2.2, 0.98\}$. For any shear parameter in $\tilde{\mathcal D}$, we use its nearest integer to denote the beam-shear mode, \emph{e.g.} $\tilde d = 12$ beam-shear mode denotes the separation distance between optically sheared beams is 5.30 $\mu$m.

\begin{figure}
\centering
\includegraphics[width=0.45\textwidth]{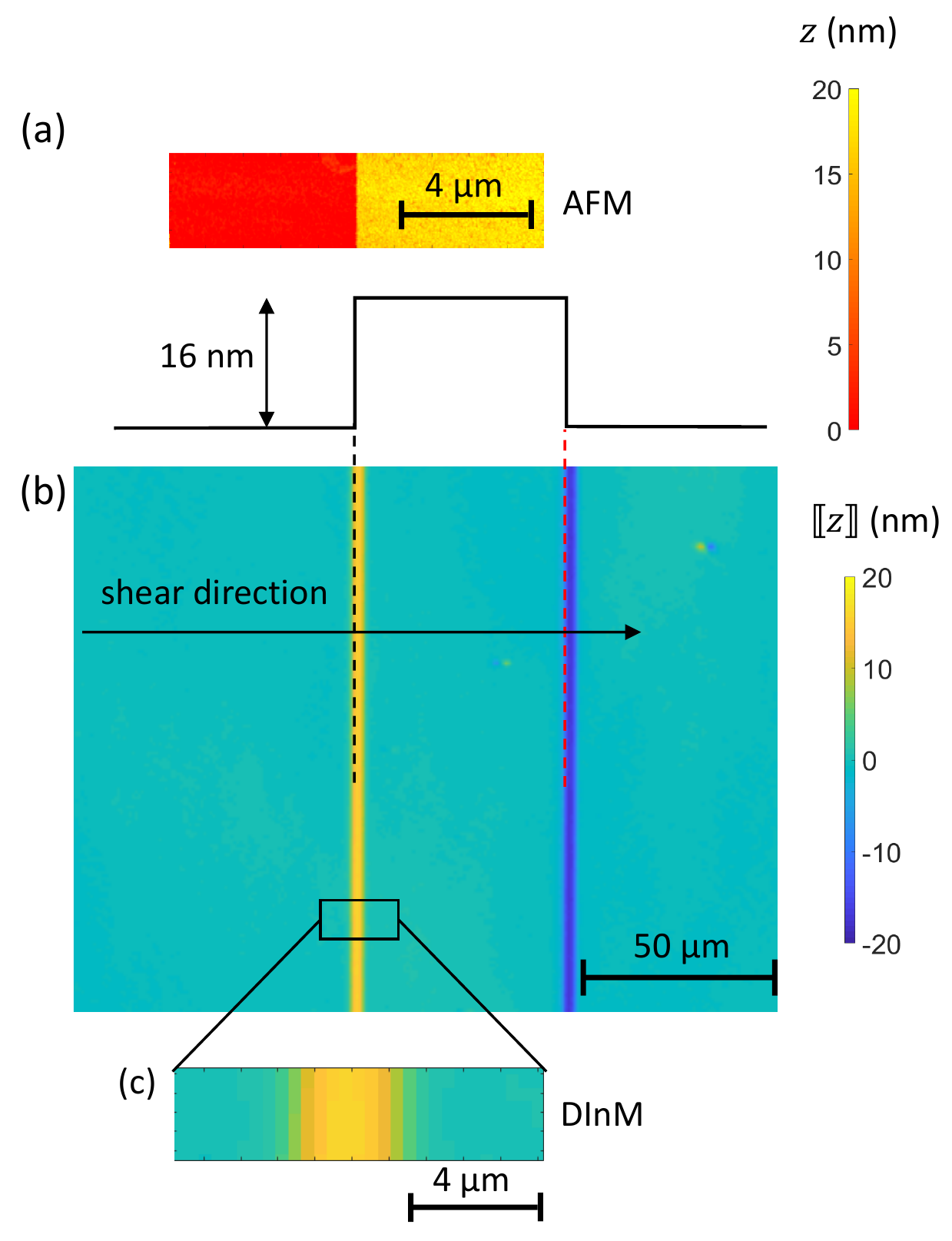}
\caption{Demonstration of nanometer step height characterization by differential interference microscopy (DInM). (a) The surface topography measurement conducted by atomic force microscope (AFM) as a reference. (b) The DInM image covering the same step, color mapped by the height variation $\llbracket z\rrbracket$, corresponding to (c) the close-up region in the vicinity of the step. }\label{fig:image}
\end{figure}

To consistently verify the surface steps characterized by DInM, we use the AFM (Digital Instruments, D3100, 0.1 nm accuracy) as the reference measurement to characterize all fabricated steps prior to the DInM measurements. The pre-scanned steps by AFM are used to calculate the theoretical $\hat{\mathcal I}^*$ by Eq.~\eqref{eq:convolution} and corresponding $\llbracket z^*\rrbracket$ by Eq.~\eqref{eq:zjump}. We also use the step height determined by AFM as a reference to evaluate the accuracy of the height variation by Eq.~\eqref{eq:epsilon} at different shear parameters. 

\begin{figure*}
\centering
\includegraphics[width=0.9\textwidth]{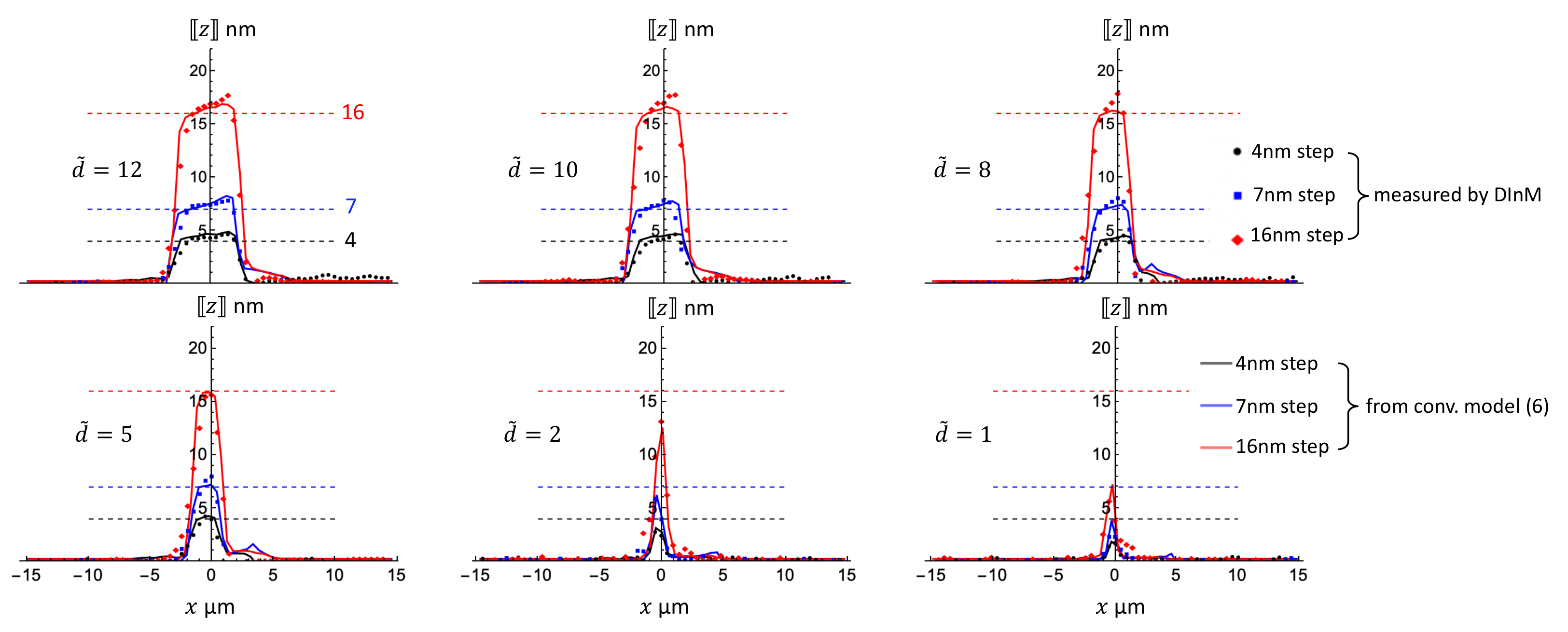}
\caption{The surface height variations characterized by differential interference microscope (DInM) for fabricated steps with nominal height $h=4, 7, 16$ nm at various shear parameters $\tilde d$ varying from 1 to 12. The dashed lines suggest the fabricated step depths. The experiment results are directly compared to the theoretical model given by Eq. \eqref{eq:convolution} without any fitting parameter. }\label{fig:meas}
\end{figure*}

Figure \ref{fig:image} displays the results of the surface step with designed depth of 16 nm, characterized by both AFM and DInM respectively. The horizontal axis ($x$-axis) of the DInM image is the beam-shear direction. Within the spatial range of 10 $\mu$m across the step, the reference measurement by AFM gives the surface height distribution with the step height $h = 16.19$ nm. 
Within the field of view that consists of the same step, we take three independent images by DInM at the different bias phases by tuning the LC retarder (Thorlabs LCC1221-A), that is $\mathcal I|_{\phi_0 = 0}, \mathcal I|_{\phi_0 = \frac{\pi}{4}}$, and $\mathcal I|_{\phi_0 = \frac{\pi}{2}}$. By Eqs. \eqref{eq:intensity} and \eqref{eq:normalizedintensity}, we have
\begin{eqnarray}
&\dfrac{\mathcal I|_{\phi_0 = 0} - \Is}{\mathcal I|_{\phi_0 = \pi/4} - \Is} = \dfrac{2 \sin^2(k \zjump)}{1 + \sin (2 k \zjump)}, \label{eq:var1}\\ 
&\dfrac{\mathcal I|_{\phi_0 = 0} - \Is}{\mathcal I|_{\phi_0 = \pi/2} - \Is} = \tan^2 (k \zjump).\label{eq:var2}
\end{eqnarray}
By the set of two equations, we can solve the surface height variation $\zjump$ by eliminating the stray light intensity $\Is$. Figure \ref{fig:image}(b) shows the two-dimensional graph of the step height variation by solving Eqs. \eqref{eq:var1} and \eqref{eq:var2}. The beam shear direction is perpendicular to the step corresponding to a positive height variation for an increase of surface height, vice versa. Figure~\ref{fig:image}(c) is the blow-up of the step corresponding to the corresponding region characterized by AFM. The broadening of the signal is due to the light differentiation, given by Eq. \eqref{eq:var}. The width of it equals to the beam-shear distance, while the blurriness at edges reveals the diffraction effect of light. The measured height variation is $\zjump = 16.69$ nm by DInM, which deviates from the AFM measurement (16.19 nm) by 0.5 nm. This demonstration validates our method at nano scales for the $> 100$ $\mu{\rm m}^2$ area at a much faster speed.

Figure \ref{fig:meas} presents the one-dimensional surface height variations for fabricated steps with designed depths of 4nm, 7nm and 16nm, characterized by our DInM system at different shear parameters in the set $\tilde{\mathcal D}$. Let $x$-axis (beam-shear direction) be aligned with the variation direction of the step. The $y$-axis is perpendicular to it, along which no height variation is observed. The 1D profile is computed as the algebraic average of the height variations as 
\begin{equation}\label{eq:zmeas}
    \left<\zjump \right> (x_i) = \frac{1}{H}\sum_{j \in [0, H]} \zjump (x_i, y_j),
\end{equation}
where $H$ denotes the number of pixels along $y$-axis of the image.  Within a spatial range from -15 $\mu$m to 15 $\mu$m that fully covers the step, the value of $\zjump$ for each of the pixels is directly calculated by Eq.~\eqref{eq:zmeas} based on the intensity profile of the DInM image. As shown in Fig. \ref{fig:meas}, the measured step profile by DInM varies as the shear parameter $\tilde d$. When $\tilde d$ is small, \emph{i.e.} $\tilde d < 5$, the measured step profile substantially deviates from its reference profile. As increasing of $\tilde d$, all three measured steps converge to their reference profiles. In particular, for $\tilde d = 12$, the DInM can reveal the step variation with sufficient axial resolution in nanometer scales.    

\begin{figure}
    \centering
    \includegraphics[width=0.35\textwidth]{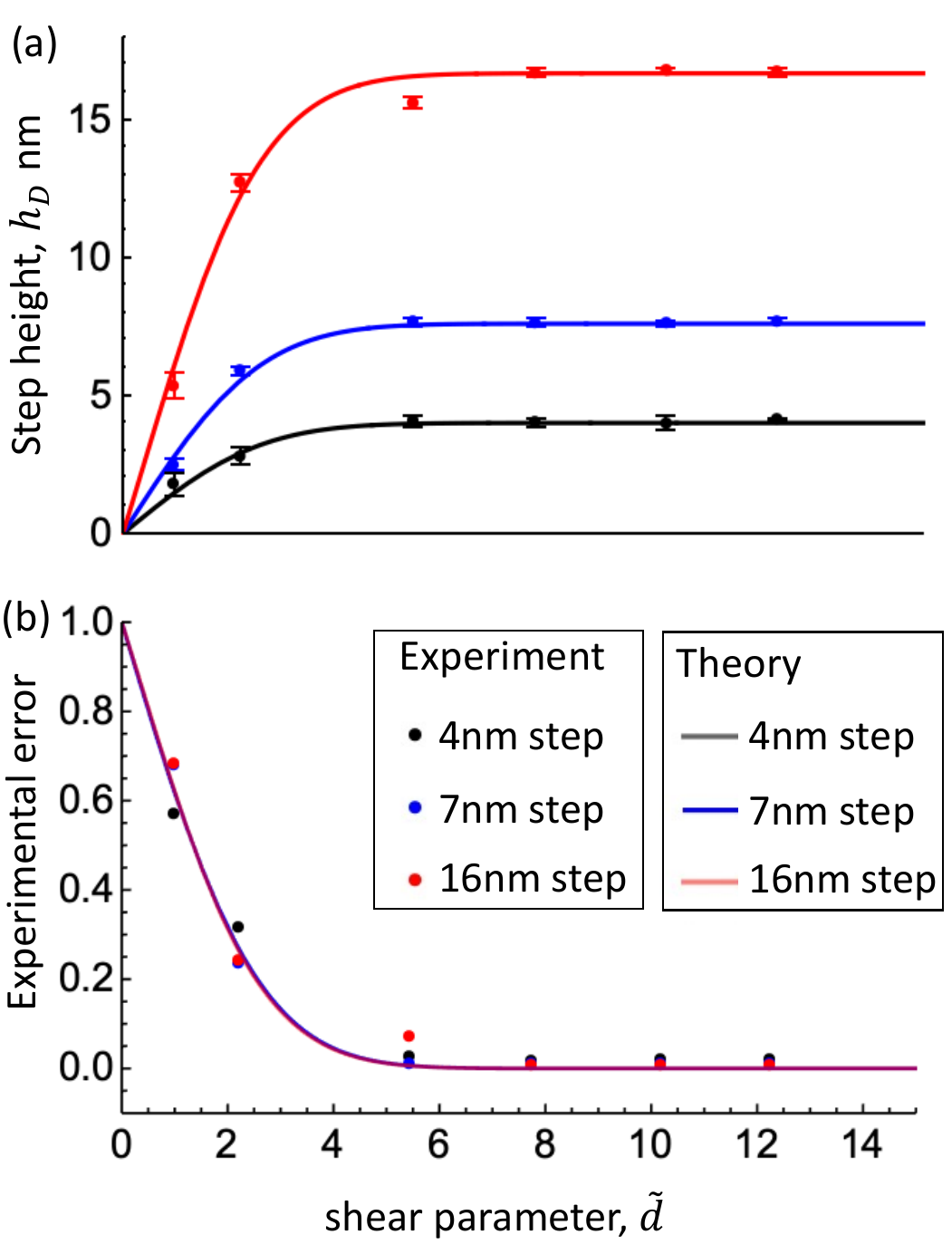}
    \caption{(a) Quantitative analysis of step heights, $h_D$ for the fabricated nanometer steps by DInM. (b) Experimental error evaluated as the discrepancy between $h_D$ and the reference height characterized by AFM. As a comparison, the solid lines are theoretical step height and corresponding diffraction-induced error directly calculated from the convolution model with different shear parameters. The error bar denotes the standard deviation over 30 DInM measurements.}
    \label{fig:h}
\end{figure}

We quantitatively analyze the accuracy of step height determined by DInM, with respect to the reference measurement by AFM. Let $\tilde x_D$ be the normalized position of the step edge. For a shear parameter $\tilde d \in \tilde{\mathcal D}$, the depth of a surface step is calculated as 
\begin{equation}
    h_D = \left<\zjump(\tilde x_D < \tilde x < \tilde x_D + \tilde d)\right> - \left<\zjump(\tilde x < \tilde x_D)\right>,
\end{equation}
where the symbol $\left<\cdot\right>$ denotes the algebraic average of $\zjump$ over all $\tilde x$ within the normalized spatial range. Figure \ref{fig:h}(a) shows the nanometer steps characterized by DInM with respect to different shear parameters. As seen, for $\tilde d > 5$, the height of each of three steps converges to $h_D = 16.69, 7.62, 4.02$ nm. The corresponding step heights determined by AFM are $16.19, 7.23,  4.24$ nm. For each of the shear parameters, we conduct 30 DInM measurements and calculate the standard deviation around the mean value of $h_D$, shown as the error bars in Fig. \ref{fig:h} (a). We observe that the variance of the measured step height gets smaller as the shear parameter goes larger. It means that the accuracy of the surface characterization can be definitively improved by using the differential interference microscope with a large beam-shear distance. The experimental error of $h_D$ is defined as
\begin{equation}
    {\rm error} = \left \vert \frac{h_D}{h_A} - 1 \right \vert,
\end{equation}
where $h_A$ is the step height measured by AFM. The experimental error that arises in the DInM measurement is plotted in Fig. \ref{fig:h}(b). As expected, the error is big for the small shear parameter, which asymptotically reduces to zero as increasing the shear parameter. We plot the diffraction-induced errors for these three nanometer steps by Eqs. \eqref{eq:zjump} and \eqref{eq:epsilon}. The trend of diffraction-induced errors well agree with the experimental errors. It indicates that the diffraction effect is the leading factor that hinders the accuracy of surface characterization by DInM optical system.  

\begin{figure}
    \centering
    \includegraphics[width=0.42\textwidth]{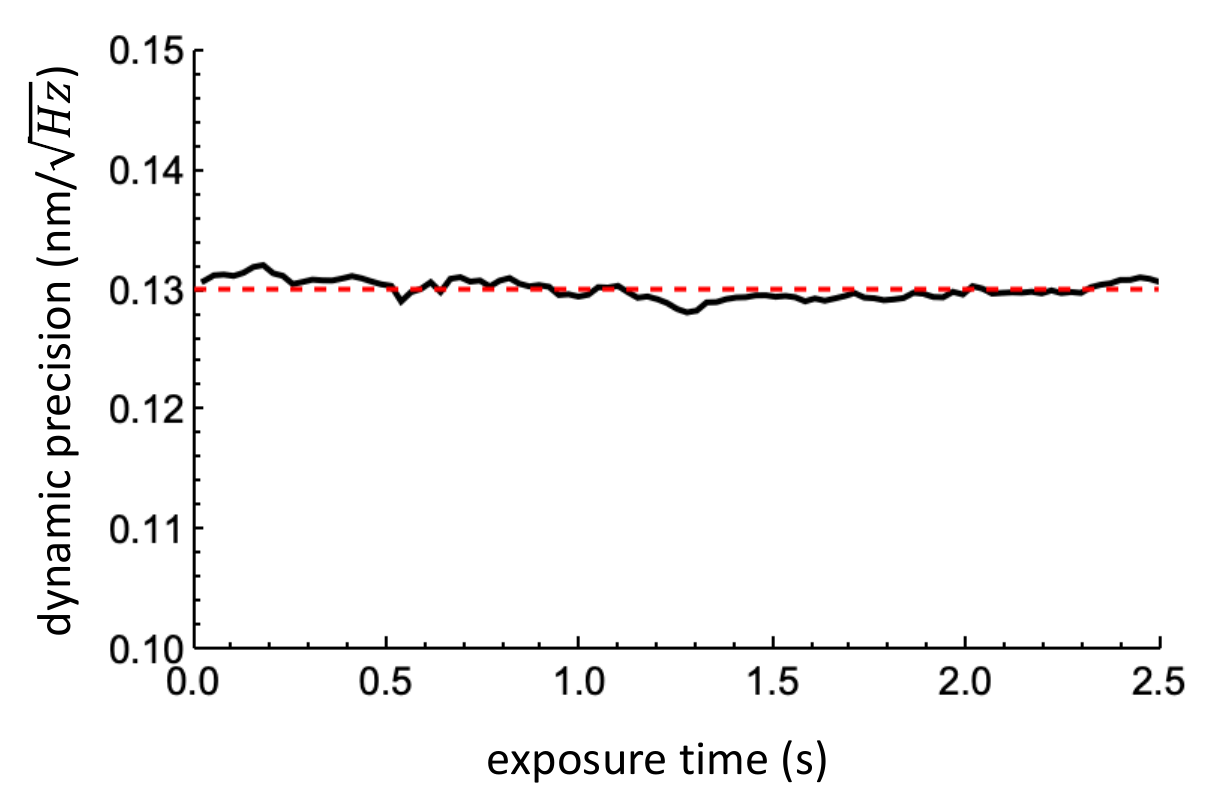}
    \caption{The dynamic measurement precision of our DInM as a function of the exposure time. The dashed line denotes the mean value of the measured precision of 0.13 nm$/\sqrt{\mathrm{Hz}}$.}
    \label{fig:derr}
\end{figure}

When considering the dynamic surface characterization by DInM, there may exist a random error caused by signal-to-noise (SNR) ratio in addition to the diffraction-induced error, discussed in the end of section \ref{sec:theory}. This random error follows Poisson distribution. As real-time measurement is employed, the SNR is improved and the random error is reduced for a longer exposure time per frame or a lower frame rate, with a cost of slower speed. On the other side, a higher frame rate with shorter exposure time leads to reduced SNR and larger measurement random error. For our DInM optical system, it takes $T = 25.5$ ms for each of the effective data points. The random error is measured as low as $\delta_T = 0.8$ nm, which corresponds to a measurement precision $\delta_T \sqrt{T} = 0.13$ nm$/\sqrt{\mathrm{Hz}}$. This is further confirmed by the measurement precision $\delta_T \sqrt{T}$ as a function of exposure time $T$ in Fig.~\ref{fig:derr}. 

\section{Conclusion}
In summary, we demonstrate subnanometer accuracy of step height measurement by the reflected-light DInM optical system with a measurement precision 0.13~nm$/\sqrt{\mathrm{Hz}}$. We propose an analytical model for step height determination, which reveals that the accuracy and experimental error are sensitive to the shear parameter $\tilde d = d/\sigma$. We conclusively show that the axial surface height variation can be characterized for nanometer steps with subnanometer accuracy. Since the optical system provides the fast imaging with large field of view, this paper opens a new avenue to make the traditional differential interference contrast microscope quantitative and accurate for surface characterization.

% \end{linenumbers}

\appendix
\section{Diffraction Effect of a DInM Image with Spatially Incoherent Light Illumination} \label{A1}

Here we provide the validation for Eq. \eqref{eq:convolution} in the 1D case. 
The uniform LED light illumination is treated as spatially completely incoherent, that is the amplitude $E_0$ in transverse plane is uniform, but the phase distribution $\varphi(x)$ is completely stochastic and uncorrected at position $x$. The spatial phase modulation caused by the light path differentiation is converted into intensity modulation through interference. The electric field of the DInM image (light-reflected mode) without diffraction effect is expressed as
\begin{equation}
E(x)=E_0 e^{i\left[2 k z(x)+\varphi(x) + 2 \phi_0\right]} - E_0 e^{i\left[2 k z(x-d) + \varphi(x)\right]},
\label{eq:Ex}
\end{equation}
where $E_0 \in \mathbb R$ is the amplitude, $\varphi(x)$ is the stochastic and uncorrelated phase fluctuation, $\phi_0$ is the bias phase tuned by the phase tuning module, and $z(x)$ is the surface height variation function.The intensity distribution in absence of the diffraction effect is calculated as
\begin{equation}
\mathcal I(x)=\vert E(x)\vert^2=4\vert E_0\vert^2  \sin^2 \left( k\zjump+\phi_0 \right).
         \label{eq:Ix}
\end{equation}
Let $H(x)$ be the optical transfer function (OTF) caused by finite aperture. The electric field on the imaging plane is obtained by
\begin{eqnarray}
E_D (x) &=& E(x) \circledast H(x) \\ \nonumber
% &=&E_0 e^{i[kz(x)+\varphi(x)+\phi_0]}\otimes h(x)+E_0 e^{i[kz(x-d)+\varphi(x)]}\otimes h(x) \\ \nonumber
            %  &=&[E_0 e^{i[kz(x)+\varphi_0]}+E_0 e^{ikz(x-d)} ] e^{i\varphi(x)} \otimes h(x)	 \\ \nonumber
             &=& E_0 f(x)e^{i\varphi(x)} \circledast H(x),			
  \label{eq:EDx}
\end{eqnarray}
where  
\begin{equation} \label{eq:fx}
f(x)=e^{i[kz(x)+\phi_0]}+e^{ikz(x-d)}.
\end{equation}
Consequently, the intensity distribution of the image becomes
\begin{equation}
\mathcal I^* (x)=\langle\vert E_D (x)\vert^2 \rangle_{\{\varphi\}},
         \label{eq:I1x}
\end{equation}
where $\langle \cdot \rangle_{\{\varphi\}}$ is the statistical average over the random and stochastic phase fluctuation $\{\varphi(x)\}$. Equation \eqref{eq:I1x}  is computed as
\begin{eqnarray}
\mathcal I^* (x)&=&\langle |E_0 f(x)e^{i\varphi(x)} \circledast H(x)|^2 \rangle_{\{\varphi\}} \\ \nonumber
           &=&|E_0|^2 \int \int H(x-u) H^*(x-v) f(u) f^*(v)\\ \nonumber
           &\times&\langle e^{i[\varphi(u)-\varphi(v)]}\rangle_{\{\varphi\}} dudv.
   \label{eq:I2x}
\end{eqnarray}
For completely spatial incoherent light beam (or the spatial coherence is limited within one wavelength $\lambda$), we have \cite{StatisticalOptics}
\begin{equation}
 \langle e^{i[\varphi(u)-\varphi(v)]}\rangle_{\{\varphi\}}\cong \lambda \delta(u-v).
 \label{eq:phase}
\end{equation}
Substituting Eq. \eqref{eq:phase} to Eq. (A6), we obtain 
\begin{equation}
\mathcal I^* (x)
           =|E_0 f(x)|^2 \circledast \lambda|H(x)|^2
           =\mathcal I(x) \circledast {\rm PSF}(x),
\label{eq:I3}           					%(S8)
\end{equation}
where the intensity point spread function (PSF) takes square of the OTF of the imaging system
\begin{equation}
	{\rm PSF}(x)=\lambda|H(x)|^2.						%(S11)
\label{eq:PSFA}           					%(S8)
\end{equation}
After normalization, Eq. \eqref{eq:I3} becomes
\begin{equation}
\mathcal \image^*(x)=\mathcal \image(x) \circledast {\rm PSF}(x).
\label{eq:I3nomarlized}           					%(S8)
\end{equation}
Note that Eq. \eqref{eq:convolution}  in main manuscript is the same as Eq. \eqref{eq:I3nomarlized}  here. For more general discussion of incoherent optical imaging, one can refer to Goodman's Introduction to Fourier Optics \cite{fourieroptics}. Here we have neglected the partial spatial coherence induced by the aperture of the objective lens. A more accurate but complicated model taking into account partial spatial coherence can follow the treatment in Ref. \cite{MehtaOE2008}.  

\section{Point Spread Function (PSF)} \label{A2}

According to Fraunhofer diffraction theory, the PSF for an ideal optical system is an Airy function. Here we show that we can use a Gaussian function to approximate the PSF for our imaging system with a sufficient accuracy. The benefit of Gaussian function is to derive analytical expression for Eq.~\eqref{eq:I3}.

In a 2D case, the Airy disk pattern of aperture-induced intensity PSF is given by
\begin{equation}
	P(\rho)=\big|\frac{J_1 ({\rm NA}k\rho)}{\rho}\big|^2,						%(S12)
\label{eq:Airy}           				
\end{equation}
where $\rho$ is the radial position in the polar coordinate, ${\rm NA}$ is the numerical aperture, $J_1$ is the order 1 Bessel function of the first kind. This Airy disk pattern radial distribution can be approximated by
\begin{equation}
G(\rho)=e^{-\rho^2/(2\sigma^2 )}.					%(S13)
\label{eq:GA}           				
\end{equation}
With $\sigma = 1.3/({\rm NA} k)$, the likeness between Eqs. \eqref{eq:Airy} and \eqref{eq:GA} is computed as
\begin{equation}
	\frac{|\int P(\rho) G(\rho) 2\pi\rho d \rho|^2}{\int|P(\rho)|^2 2\pi\rho d\rho\times \int|G(\rho)|^2 2\pi\rho d\rho}=0.9954.			%(S14)
	\label{eq:likeness1}           				
\end{equation}
It shows that the Airy disk pattern can be well approximated by the Gaussian function. 
% In Fig. \ref{fig:PSF}(a), we plot the comparison between both functions.

In 1D case, the aperture-confined PSF is a Sinc function
\begin{equation}
	P(x)=\big|\frac{\sin({\rm NA}kx)}{{\rm NA}kx}\big|^2,						%(S12)
\label{eq:sink}           				
\end{equation}				
which can be approximated by  
\begin{equation}
G(x)=e^{-x^2/(2\sigma^2 )}.					%(S13)
\label{eq:GS}           				
\end{equation}
With $\sigma = 1.1/({\rm NA} k)$, the likeness between Eqs. \eqref{eq:sink} and \eqref{eq:GS} is 
\begin{equation}
	\frac{|\int P(x) G(x) dx|^2}{\int|P(x)|^2 dx\times \int|G(x)|^2 dx}=0.9956.	
	\label{eq:likeness2}           				
\end{equation}
It also shows that the Gaussian function is a valid approximation of the Sinc function by a proper selection of $\sigma$.
% In Fig. \ref{fig:PSF}(b), we plot both distributions for a comparison. 

% Therefore, in this work, we choose Gaussian function for the PSF which allows us to derive the analytic expression of Eqs. \eqref{eq:normalizedintensity2} and \eqref{eq:errorfunc}. Our experimental results also confirm the Gaussian function describes well the PSF of our DInM system. 

\section*{Conflict of interests}
The authors declare no financial/commercial Conflict of Interest.

\section*{Acknowledgments}

K. H. C. and X. C. are grateful for financial support under GRF grants 16201019 and 16203021, CRF grant C6016-20G from the HK Research Grants Council. 

\bibliography{ZResolution}

\end{document}